\documentclass[twoside,twocolumn,9pt]{article}
\usepackage{extsizes}
\usepackage[super,sort&compress,comma]{natbib} 
\usepackage[version=3]{mhchem}
\usepackage[left=1.5cm, right=1.5cm, top=1.785cm, bottom=2.0cm]{geometry}
\usepackage{balance}
\usepackage{mathptmx}
\usepackage{sectsty}
\usepackage{graphicx} 
\usepackage{lastpage}
\usepackage[format=plain,justification=justified,singlelinecheck=false,font={stretch=1.125,small,sf},labelfont=bf,labelsep=space]{caption}
\usepackage{float}
\usepackage{fancyhdr}
\usepackage{fnpos}
\usepackage[english]{babel}
\addto{\captionsenglish}{%
  
}
\usepackage{array}
\usepackage{droidsans}
\usepackage{charter}
\usepackage[T1]{fontenc}
\usepackage[usenames,dvipsnames]{xcolor}
\usepackage{setspace}
\usepackage[compact]{titlesec}
\usepackage{hyperref}

\usepackage{caption}
\usepackage{subcaption}
\usepackage{epstopdf}

\definecolor{cream}{RGB}{222,217,201}

\begin{document}

\pagestyle{fancy}
\thispagestyle{plain}
\fancypagestyle{plain}{
\renewcommand{\headrulewidth}{0pt}
}

\makeFNbottom
\makeatletter
\renewcommand\LARGE{\@setfontsize\LARGE{15pt}{17}}
\renewcommand\Large{\@setfontsize\Large{12pt}{14}}
\renewcommand\large{\@setfontsize\large{10pt}{12}}
\renewcommand\footnotesize{\@setfontsize\footnotesize{7pt}{10}}
\makeatother

\renewcommand{\thefootnote}{\fnsymbol{footnote}}
\renewcommand\footnoterule{\vspace*{1pt}%
\color{cream}\hrule width 3.5in height 0.4pt \color{black}\vspace*{5pt}} 
\setcounter{secnumdepth}{5}

\makeatletter 
\renewcommand\@biblabel[1]{#1}            
\renewcommand\@makefntext[1]%
{\noindent\makebox[0pt][r]{\@thefnmark\,}#1}
\makeatother 
\renewcommand{\figurename}{\small{Fig.}~}
\sectionfont{\sffamily\Large}
\subsectionfont{\normalsize}
\subsubsectionfont{\bf}
\setstretch{1.125} 
\setlength{\skip\footins}{0.8cm}
\setlength{\footnotesep}{0.25cm}
\setlength{\jot}{10pt}
\titlespacing*{\section}{0pt}{4pt}{4pt}
\titlespacing*{\subsection}{0pt}{15pt}{1pt}

\fancyfoot{}
\fancyfoot[LO,RE]{\vspace{-7.1pt}\includegraphics[height=9pt]{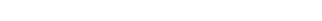}}
\fancyfoot[CO]{\vspace{-7.1pt}\hspace{13.2cm}\includegraphics{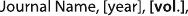}}
\fancyfoot[CE]{\vspace{-7.2pt}\hspace{-14.2cm}\includegraphics{head_foot/RF}}
\fancyfoot[RO]{\footnotesize{\sffamily{1--\pageref{LastPage} ~\textbar  \hspace{2pt}\thepage}}}
\fancyfoot[LE]{\footnotesize{\sffamily{\thepage~\textbar\hspace{3.45cm} 1--\pageref{LastPage}}}}
\fancyhead{}
\renewcommand{\headrulewidth}{0pt} 
\renewcommand{\footrulewidth}{0pt}
\setlength{\arrayrulewidth}{1pt}
\setlength{\columnsep}{6.5mm}
\setlength\bibsep{1pt}

\makeatletter 
\newlength{\figrulesep} 
\setlength{\figrulesep}{0.5\textfloatsep} 

\newcommand{\topfigrule}{\vspace*{-1pt}%
\noindent{\color{cream}\rule[-\figrulesep]{\columnwidth}{1.5pt}} }

\newcommand{\botfigrule}{\vspace*{-2pt}%
\noindent{\color{cream}\rule[\figrulesep]{\columnwidth}{1.5pt}} }

\newcommand{\dblfigrule}{\vspace*{-1pt}%
\noindent{\color{cream}\rule[-\figrulesep]{\textwidth}{1.5pt}} }

\makeatother

\twocolumn[
  \begin{@twocolumnfalse}
{\includegraphics[height=30pt]{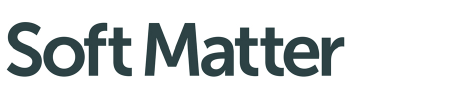}\hfill\raisebox{0pt}[0pt][0pt]{\includegraphics[height=55pt]{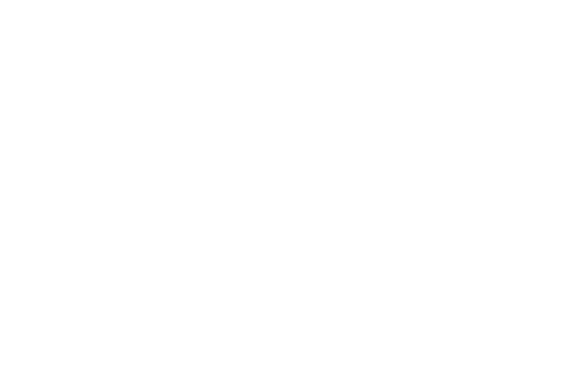}}\\[1ex]
\includegraphics[width=18.5cm]{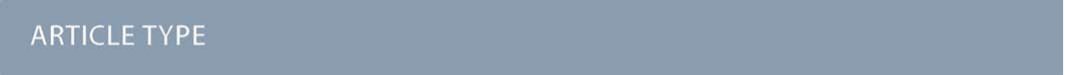}}\par
\vspace{1em}
\sffamily
\begin{tabular}{m{4.5cm} p{13.5cm} }

\includegraphics{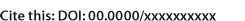} & \noindent\LARGE{\textbf{Controlling the size and adhesion of DNA droplets using surface-active DNA molecules}} \\
\vspace{0.3cm} & \vspace{0.3cm} \\

 & \noindent\large{Daqian Gao,\textit{$^{a}$} Sam Wilken,\textit{$^{a,b}$} Anna Nguyen,\textit{$^{c}$} Gabrielle R. Abraham,\textit{$^{a}$} Tim Liedl,\textit{$^{d}$} and Omar A. Saleh\textit{$^{a,c,e}$}} \\

\includegraphics{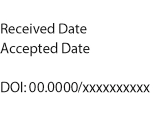} & \noindent\normalsize{Liquid droplets of biomolecules serve as organizers of the cellular interior and are of interest in biosensing and biomaterials applications. Here, we investigate means to tune the interfacial properties of a model biomolecular liquid consisting of multi-armed DNA 'nanostar' particles. We find that long DNA molecules that have binding affinity for the nanostars are preferentially enriched on the interface of nanostar droplets, thus acting as surfactants. Fluorescent measurements indicate that, in certain conditions, the interfacial density of the surfactant  is around 20 per square micron, indicative of a sparse brush-like structure of the long, polymeric DNA. Increasing surfactant concentration leads to decreased droplet size, down to the sub-micron scale, consistent with arrest of droplet coalescence by the disjoining pressure created by the brush-like surfactant layer. Added DNA surfactant also keeps droplets from adhering to both hydrophobic and hydrophilic solid surfaces, apparently due to this same disjoining effect of the surfactant layer. We thus demonstrate control of the size and adhesive properties of droplets of a biomolecular liquid, with implications for basic biophysical understanding of such droplets, as well as for their applied use.}

\end{tabular}

 \end{@twocolumnfalse} \vspace{0.6cm}

  ]

\renewcommand*\rmdefault{bch}\normalfont\upshape
\rmfamily
\section*{}
\vspace{-1cm}


\footnotetext{\textit{$^{a}$~Physics Department, University of California, Santa Barbara, California 93106, USA }}
\footnotetext{\textit{$^{b}$~Institute of Collaborative Biotechnologies, University of California, Santa Barbara, California 93106, USA }}
\footnotetext{\textit{$^{c}$~Biomolecular Science \& Engineering Program, University of California, Santa Barbara, California 93106, USA }}
\footnotetext{\textit{$^{d}$~Department für Physik, Ludwig-Maximilians-Universität, München 80539, Germany }}
\footnotetext{\textit{$^{e}$~Materials Department, University of California, Santa Barbara, California 93106, USA. E-mail: saleh@engineering.ucsb.edu}}


Biomolecular liquid-liquid phase separation plays an important role in the function of living cells. The formation of biomolecular droplets is thought to spatially organize the cellular interior by augmenting or depleting the concentration of various components, enabling control of biochemical reactions,  which are enhanced by the dynamic, liquid-like behavior of the host droplet \cite{Shin2017, Banani2017, Hyman2014}. Inspired by such properties, various works have studied the \emph{in vitro} creation of biomolecular droplets with similar properties \cite{Yewdall2021,Martin2019}, with the goals of creating artificial bioreactors and/or synthetic cells that offer novel chemical or therapeutic functionalities, or of carrying out high-resolution  measurements that probe the fundamental physical properties of phase-separated biomolecular systems. 

Spontaneous phase separation is a stochastic, thermally-driven process which leads to polydisperse distributions of droplet sizes \cite{Binder2001}. Such distributions might not be ideal for certain applications. For example, monodisperse droplets can pack more densely ($\sim 74\%$ for crystalline arrangements) than marginally polydisperse distributions ($\sim 64~\%$ for $10~\%$ polydispersity\cite{desmond2014}). Further, the timescale of solute transport within a droplet will vary with droplet size, suggesting that monodisperse droplets will behave more consistently as biosensors\cite{gong2022}. Therefore, studies of droplet size regulation are needed, and may also reveal physical mechanisms relevant to biological condensates.  

Here, we explore size control of biomolecular droplets using a model phase-separating system consisting of multi-armed DNA particles, termed nanostars (NSs) \cite{Biffi2013}. We use 4-armed NSs that are created through sequence-specific self-assembly, with each arm carrying a palindromic sticky end. Below a critical temperature, the sticky ends hybridize which, along with the multivalent nature of the particles, leads to the formation of dense, disordered, and dynamic DNA meshworks that exhibit bulk liquid-like properties~\cite{Jeon2018}. Prior work demonstrates that this is an equilibrium liquid-liquid phase separation process \cite{Rovigatti2014pd, Rovigatti2014, Conrad2022} with a well-defined regime at intermediate NS concentrations at which micron-scale liquid DNA droplets can coexist with a dilute nanostar solution.

Inspired by the principles used in emulsion stabilization and dispersion polymerization \cite{Beck2020}, we investigate the control of NS droplet size using interfacially-active molecules. Particularly, a previous study by Nguyen \textit{et al.} \cite{Nguyen2019} showed that relatively long double-stranded DNA molecules carrying one or two sticky ends can localize to the NS droplet interface. Such `DNA surfactants' are attracted to the droplets by the sticky-end hybridization energy, but can be excluded from the droplet interior due to the entropic cost of confining long DNA strands inside the dense nanostar mesh, and so become enriched at the interface.

We find that as we increase the ratio of DNA surfactant to NSs, the mean size of droplets decreases until, ultimately, sub-micron droplets are formed that we interpret as swollen-micelle-like structures with a liquid DNA nanostar core and a corona of long DNA. We also observe that NS droplets formed in the presence of DNA surfactant resist adhering to solid surfaces, consistent with the surfactant forming a protective, brush-like layer around the droplets. Our work demonstrates an interfacial engineering strategy to achieve the combined control of condensate size and adhesive properties, which we suggest could be important in applications of biomolecular droplets.

\begin{figure}
\centering
  \includegraphics[height=9cm]{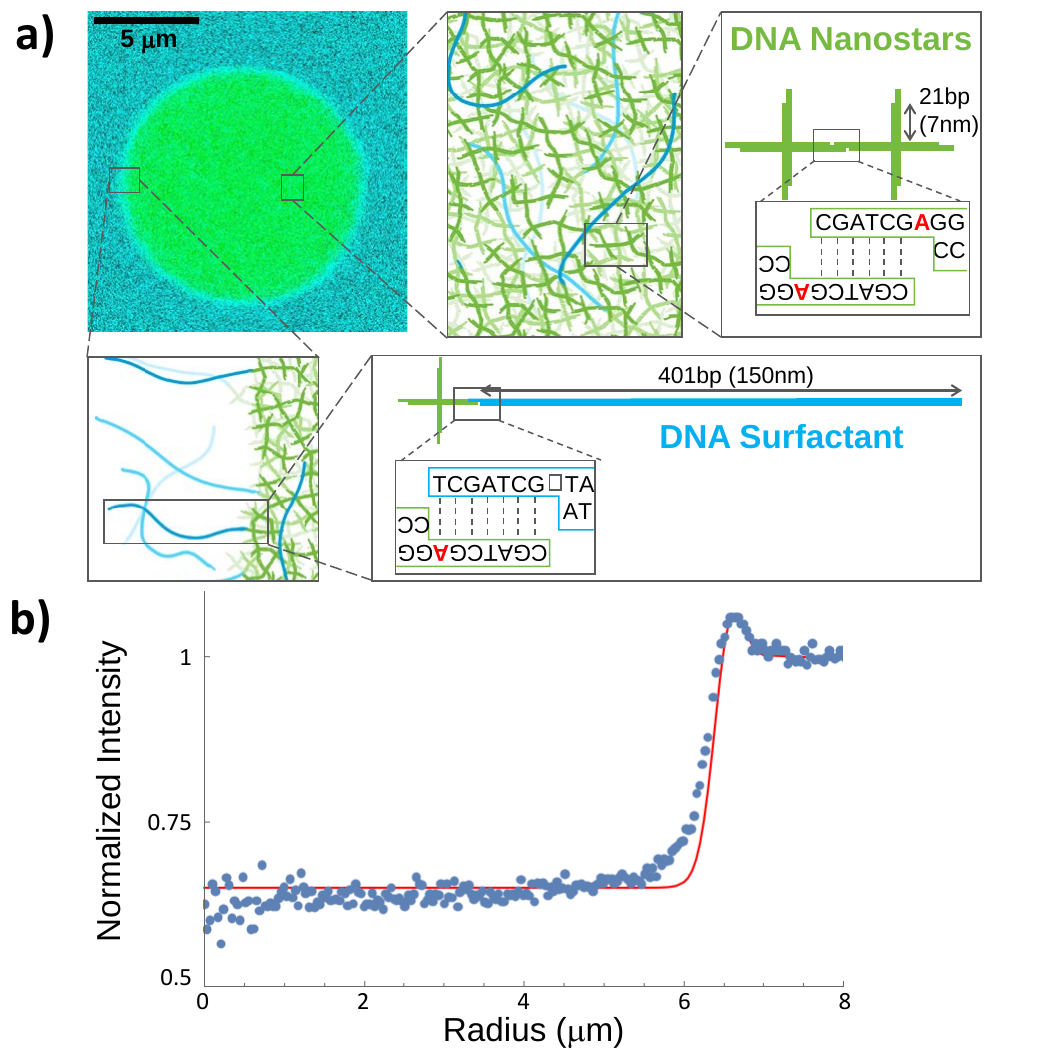}
\caption{Structural and molecular profile of the NS-surfactant system. a) Top left: A multicolor confocal fluorescent image of a surfactant coated NS droplet in a solution containing $4~\mathrm{\mu M}$ NS,  $0.4~ \mathrm{\mu M}$ DNA surfactant, and $1$ M NaCl. 100\% of the DNA surfactant molecules are fluorescently tagged with a Cy3 fluorophore, and $5\%$ of the nanostars are tagged with a Cy5 fluorophore. Cartoons depict the design of NS and surfactant molecules, and the structure of the droplet interior and interface. The depicted NS liquid structure is only schematic; prior work indicates that the bulk of the liquid is nearly fully bonded in these conditions \cite{Rovigatti2014pd}. b) The radial intensity profile of DNA surfactant fluorescence inside and outside an NS droplet. The points correspond to analysis of the image shown in (a). The line is a model of the intensity profile, adjusted to match the data,  and found by calculating the convolution of the microscope's point-spread function with a sphere (representing the droplet) coated with a bright shell (representing the surfactant coating). See main text for details. } 
\label{fig:figure1}
\end{figure}

\section*{Results and Discussions}

\subsection*{Nanostar and DNA Surfactant Design and Assembly}
We use 4-armed DNA NSs that are assembled from four single-stranded oligos, following previous designs \cite{Biffi2013,Jeon2018}. The sequence of the oligos dictates their assembly into an X-shaped structure, with four double-stranded arms extending from a common junction (Fig.~\ref{fig:figure1}). Each arm terminates in a 7 base single-stranded region consisting of a 6 base palindromic sticky end, with sequence \emph{CGATCG}, that mediates NS-NS binding, and a seventh base (Adenine) that is left unpaired upon NS-NS binding. Internal NS flexibility, due  to this unpaired base and/or the conformational freedom of the arms around the junction, is thought to influence the liquid-like behavior of the resulting condensates \cite{Rovigatti2014,Nguyen2017,Lee2021}. 

Following prior work \cite{Nguyen2019}, we designed linear DNA molecules 401~bp in length, and with a single sticky end, to act as surfactants for NS droplets. These surfactant DNA molecules were synthesized using the auto-sticky polymerase chain reaction  method\cite{Gal1999}, and contain a 7-base non-palindromic sticky end, of sequence \emph{TCGATCG} (Figure \ref{fig:figure1}). This sequence is complementary to both the palindromic and flexible base on the NSs. Thus, relative to the NS-NS binding interaction, the NS-surfactant binding interaction is significantly stronger, since it is 1 base longer and involves 2 extra base-stacking interactions (i.e. between the C and the 5' terminal T of the 7-base sequence, and between the terminal T and the last base of the NS arm). Further, the design of the 7-base sequence is intended to minimize surfactant-surfactant binding: while the 7-base surfactant sticky ends could potentially partially hybridize through the contained 6-based palindromic sequence, doing so would require an unfavorable exclusion of the terminal T base (Figure \ref{fig:figure1}).  The double-stranded portion of the surfactant (401 bp $\approx 138$~nm) is much longer than the $\approx 9$ nm mesh size estimated previously for NS condensates; thus partitioning of the surfactant into the condensate involves significant restriction of its lateral fluctuations, and is entropically unfavorable, as found previously \cite{Nguyen2019}.

\begin{figure*}
 \centering
 \includegraphics[height=5cm]{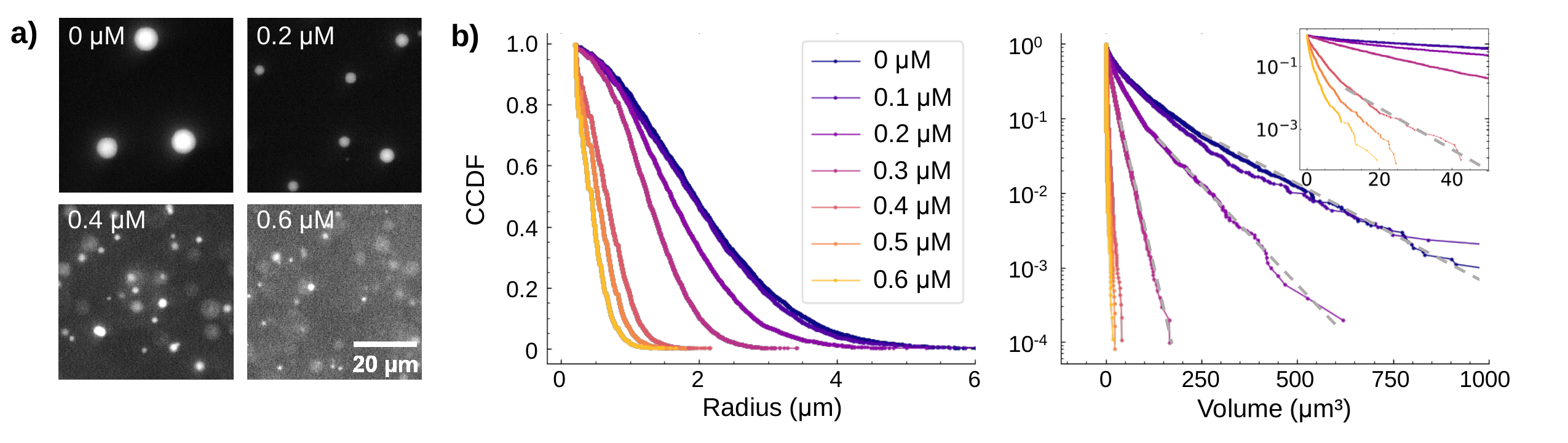}
 \caption{a) Epi-fluorescent micrographs of NS droplets in solutions containing  $2~\mu$M NS, 1 M NaCl, and the indicated concentration of DNA surfactant. b) Measured cumulative distributions of droplet sizes, as found  from images such as those shown in (a), in solutions of $2~\mu M$ NS, 1 M NaCl, and the indicated concentration of DNA surfactant. Left and right panels show the same data as a function of droplet radius (left) and of droplet volume (right; note semi-log axes). Dashed gray lines are guides to the eye, indicating exponential distributions. Inset expands the volume plot for small droplet sizes.}
 \label{fig:figure2}
\end{figure*}

Since the linear DNA constructs bind strongly to NSs, but are entropically disfavored from entering the condensate, they accumulate on the NS droplet interface (Fig.~\ref{fig:figure1}). To visualize surfactant location, we label the NSs and surfactant molecules using distinct fluorophores and image droplets using confocal microscopy. In the surfactant fluorophore channel, we observe a ring of high intensity on the droplet surface, indicating the presence of a spherical shell of surfactant surrounding the droplet, and demonstrating the surface-active nature of the construct. The images also indicate the surfactant molecules are only modestly depleted from the droplet interior (partition coefficient $\approx 0.65$); this is slightly higher than that found for DNA of similar length in a previous study\cite{Nguyen2019}, which we attribute to the stronger binding design used here, which drives higher solubility of the surfactant into the bulk droplet.

We can use the measured intensity distributions of the DNA surfactant to estimate its density on the NS droplet/dilute solution interface. For the microscope and configuration used, we independently measured the point-spread function (PSF) to have a full width at half maximum (FWHM) of roughly 1.9~$\mu$m in z, and $0.35~\mu$m in x and y. We then assumed the PSF is well-represented by a 3D gaussian with a spread, in each dimension, matching the respective FWHM value. This PSF was then convoluted with a geometric model representing the surfactant-coated NS droplet, consisting of a sphere coated with a bright shell. The resulting predicted intensity distribution was roughly fit to the data by adjusting the intensity inside and outside the sphere, the sphere radius, and the shell brightness; the result is shown in Fig.~\ref{fig:figure1}b. Knowing that the exterior solution contained 400 $\mu$M surfactant fluorophore allowed calibration of the intensity values; we thus calculated that the best-fit shell brightness corresponds to roughly 20 surfactant molecules per square micron, or, equivalently, a typical surfactant spacing of around 220 nm. This indicates that, at least for the concentrations of NS and surfactant used here, the surfactant DNA is spaced by a distance similar to its contour length (150 nm), indicating that the interface is not a true polymer brush (spacing $\ll$ polymer size), but rather a more sparse brush. This is reasonable, as polymer insertion into a true brush is energetically costly\cite{rubinstein2003}, which would not be supported by the somewhat modest surfactant/nanostar binding energy. 


\subsection*{DNA surfactants control the size of nanostar droplets} 
Given the sparse-brush geometry of the DNA surfactant, we posited that it could act as a buffer that inhibits droplet coalescence, thereby regulating the droplet size distribution.
To examine this effect, we mixed $2~ \mathrm{\mu M}$ fluorescently-labeled NSs with surfactant at a concentration varying from $0~ \mathrm{\mu M}$ to $0.6~ \mathrm{\mu M}$. The mixtures were held at 50$^{\circ}$C for $30$ minutes to melt all sticky ends, then moved to a rotor at room temperature to allow phase separation while preventing sedimentation. After $2$ hours of rotation, the droplets were transferred into BSA coated flow cells and left unperturbed for $30$ minutes to let droplets fully sediment. The droplet size distribution was then measured by fluorescent imaging.

We observed a strong decrease in NS droplet size with increasing DNA surfactant concentration (Figure \ref{fig:figure2}a). The size distribution is presented as a complimentary cumulative distribution function (CCDF), defined as $\mathrm{CCDF}=1- \int_{0}^{x} f(x')dx'$, where $f(x')$ is the probability density of observing a droplet of size $x'$. Use of the CCDF reduces noise and prevents binning effects relative to a standard histogram.

For each sample, the CCDF with respect to droplet volume is roughly consistent with an exponential distribution (Figure \ref{fig:figure2} b). Such an exponential shape is expected for a system of droplets that nucleate quickly, then coarsen and grow primarily through stochastic coalescence, as discussed recently by Lee \textit{et al.} \cite{Lee2023}. Prior work has indeed indicated that NS droplets nucleate relatively quickly at the high salt concentrations used here \cite{wilken2023}. Further, prior estimates indicate the interfacial tension of 4-armed NS droplets is extremely low, of order $1 \mu$N/m \cite{Jeon2018}, meaning the droplets have very small Laplace pressures and, accordingly, very little Ostwald ripening. This is consistent with coarsening being dominated instead by coalescence, as indicated by the shape of the CCDF \cite{Lee2023}. The effectiveness of the surfactant DNA in reducing droplet size can thus be attributed to surfactant-induced slowing of coalescence, which we attribute to a disjoining pressure created between neighboring droplets by the entropy of the sparse brush layer of surfactants.

\begin{figure}
 \centering
 \includegraphics[height=8cm]{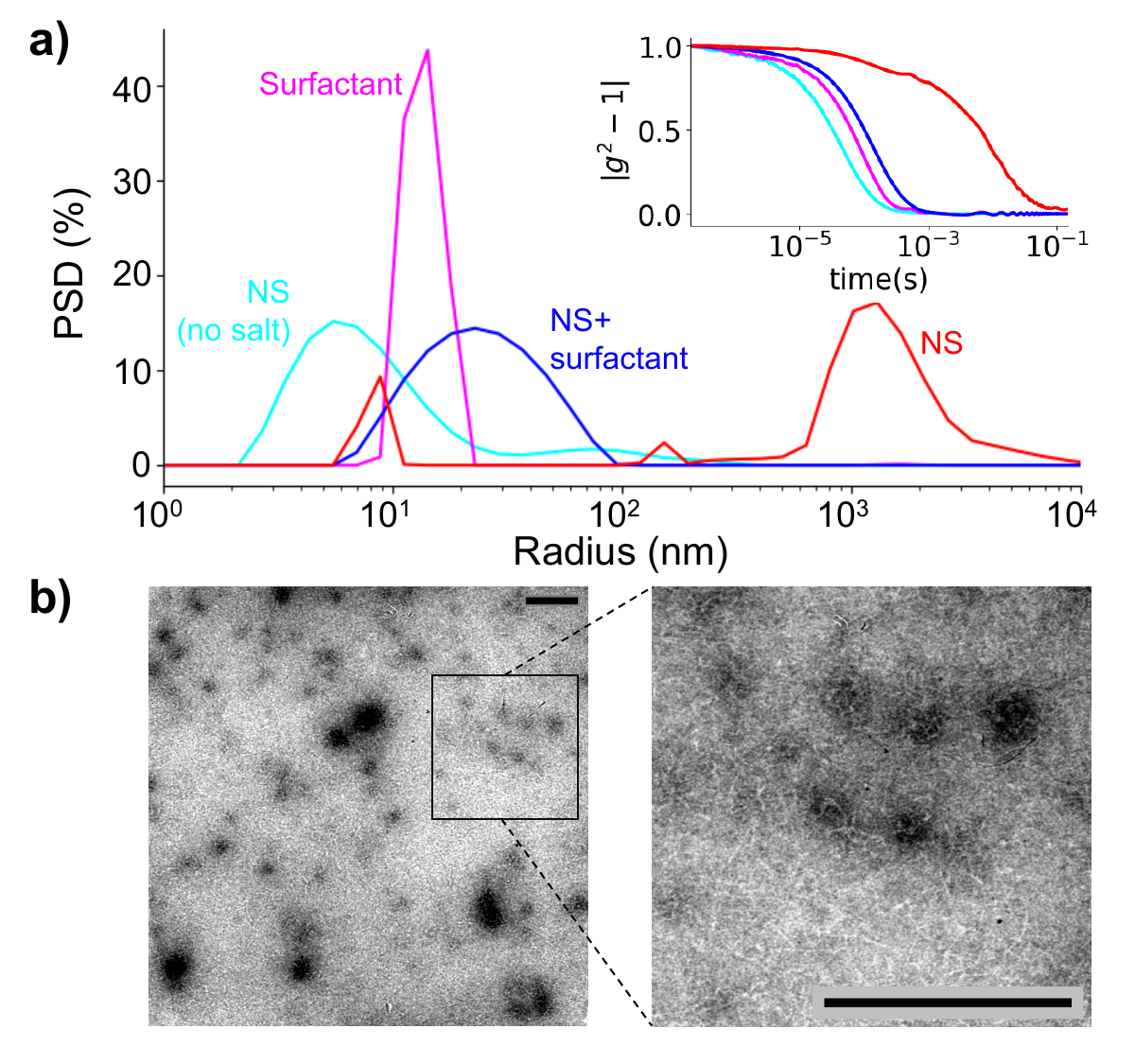}
 \caption{ a) Dynamic light scattering (DLS) measurements of particle sizes in various NS/surfactant mixtures. The inset shows the time correlation functions, and the main plot shows the distributions of hydrodynamic radii inferred from those correlation functions. Most measurements were performed in 1 M NaCl and with 2~$\mu$M NS and/or  2~$\mu$M surfactant, as indicated in the label. The 'NS (no salt)' measurement had 2~$\mu$M NS and no added salt (only buffer), a condition that disallows phase separation. All measurements were at room temperature.  b) TEM images of solutions of $0.4~ \mathrm{\mu M}$ NS and $0.4~ \mathrm{\mu M}$ DNA surfactant at $1$ M NaCl. Scale bars: $300$ nm.}
 \label{fig:figure3}
\end{figure}

\begin{figure*}
 \centering
 \includegraphics[height=4.5cm]{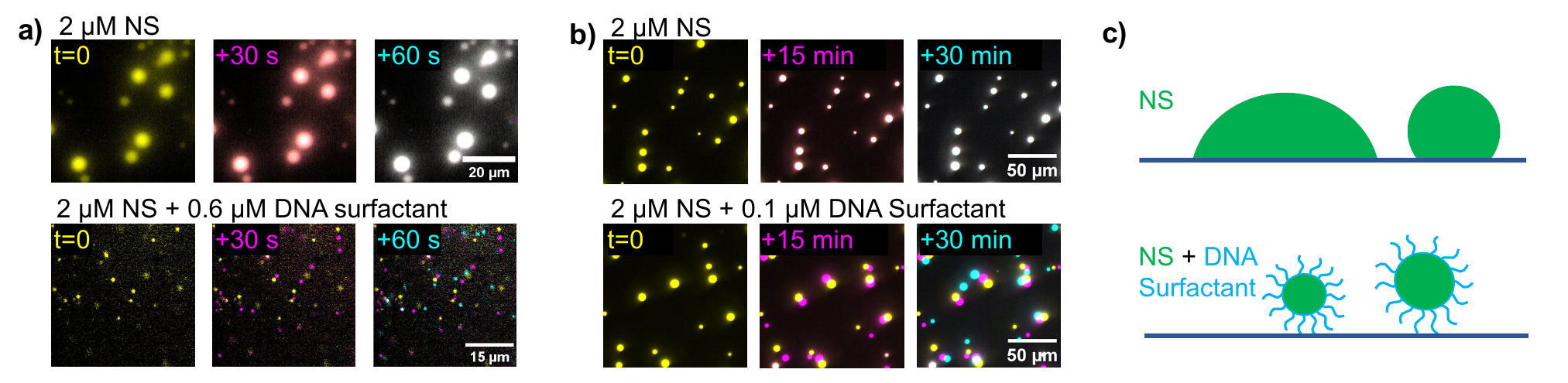}
 \caption{a) Adhesive properties of droplets near a hydrophobic surface. Each row shows confocal images of the same area at three different time points, with later-time images being superimposed on earlier-time images in a different color. NS-only droplets (top row) show color merging, indicating droplet adhesion to the surface and a lack of motion. NS droplets with DNA surfactant show significant droplet movement and a lack of adhesion. b) Adhesive properties of droplets near a hydrophilic (BSA coated) surface, as revealed by time-course confocal images, and colored as in (a). Again, NS-only droplets show adhesion, while the presence of surfactant enables droplet mobility. c) Schematic diagrams indicating droplet behavior with and without DNA surfactant coating.}
 \label{fig:figure4}
\end{figure*}

\subsection*{High surfactant to NS ratios generate sub-micron-scale structures.} 
For surfactant to NS ratios beyond $1:3$, the NS droplets grew so small as to be difficult to resolve in an optical microscope. To quantify droplets in this regime, we used dynamic light scattering (DLS). 
The scattering time correlation functions (inset, Fig.~\ref{fig:figure3}a) indicate that solutions of NSs alone have extremely slow dynamics, consistent with the presence of large, slowly-diffusing, phase-separated droplets, but that addition of surfactant greatly accelerates the dynamics. However, NS/surfactant mixtures at a $1:1$ ratio still showed dynamics slower than either surfactant alone, or NSs at high temperatures, indicating that NS/surfactant mixtures formed self-assembled structures that exceeded the size of either single NSs or single surfactant molecules. When the correlation times are interpreted as particle sizes (Fig.~\ref{fig:figure3}a), we find that the NS/surfactant solution contains a relatively broad distribution of hydrodynamic radii centered around 30 nm, which is notably larger than the hydrodynamic radii measured for either NSs alone ($\approx 7$nm) or surfactant alone ($\approx$ 10 nm). We conclude that, at these high surfactant concentrations, the solution forms nanoscale NS/surfactant assemblies. Given the surface-localized nature of the long DNA, we suggest that those assemblies are akin to swollen micelles, consisting of a core of a few tens of NSs decorated by a corona of long surfactant DNA.


To more directly probe the sub-micron NS/surfactant assemblies, we used transmission electron microscopy (TEM) to image their structure. While very high concentrations of DNA were needed to obtain a strong DLS signal, such high concentrations were not optimal for TEM imaging; thus the TEM samples were diluted down $5$ times relative to that used for DLS, with the surfactant to NS ratio unchanged. TEM imaging  involves a surface-deposition and flash-drying process that could potentially alter the structure of these non-covalent assemblies. Nonetheless, TEM images (Fig.~\ref{fig:figure3} b) reveal assemblies (dark blobs in the image) of similar size to those seen in solution by DLS, and also reveal long DNA molecules (white sinuous lines in the image) potentially indicative of a coating surfactant layer. Thus, the TEM images are consistent with the existence of swollen-micelle-like assemblies at high surfactant ratios.

\subsection*{DNA surfactants modulate the adhesive properties of nanostar droplets.} 
That the sparse brush layer of long DNA was effective in inhibiting droplet coalescence indicates it might also be effective in modulating NS droplet adhesive properties. Prior work indicated that droplets formed solely from NSs adhere to hydrophobic surfaces, with contact angles of $\theta \approx 70^\circ$ (where $\theta \rightarrow 0$ corresponds to full wetting) \cite{saleh2020}. This wetting behavior is likely caused by stacking of the single-stranded bases in the NS sticky ends against the hydrophobic surface \cite{elder2013}. Here, we replicated that prior result: we formed droplets solely from NSs, placed them in a flow cell with a hydrophobic surface, allowed them to sediment and interact with the surface for 30 minutes, then acquired a short series of time lapse images. We found that the NS-only droplets lacked Brownian motion, indicating that they stuck to the hydrophobic surface (Fig.~\ref{fig:figure4}a, top row). In contrast, NS droplets formed in the presence of the DNA surfactant, and subject to the same protocol, retained Brownian motion (Fig.~\ref{fig:figure4}a, bottom row). We conclude that the shell of long DNA around the droplets is of sufficient density and thickness to disallow adhesive contact of NS sticky ends to the hydrophobic surface (Fig.~\ref{fig:figure4}c), enabling the droplets to retain mobility.

We further found that the enhanced mobility of NS droplets in the presence of DNA surfactant occurs on other surfaces, notably a clean glass surface coated in bovine serum albumin (BSA). As indicated by time-lapse images taken with $15$ minutes interval (Fig.~\ref{fig:figure4}b, top row), NS-only droplets adhered to this surface, while adding even a relatively small amount of DNA surfactant (0.1~$\mu$M) enabled the droplets to undergo Brownian motion. This further strengthens the conclusion that the shell of DNA surfactant acts to prevent the core of the NS droplets from interacting with exterior surfaces. 

Interestingly, while $0.1 \mu$M surfactant is sufficient to passivate droplets with respect to interactions with an external solid surface, this low concentration does not have a strong effect on the droplet size (see Figs.~\ref{fig:figure2} and \ref{fig:figure4}b). Thus it seems the strength of the passivation effect of the surfactant DNA is dependent on the nature of the opposing surface, with interacting liquid surfaces requiring, apparently, higher surfactant concentrations (and presumably better surface coverage) to prevent coalescence. We speculate this might have to with the increased fluctuations of a liquid versus a solid interface, which would permit liquid droplets to make contact even at low surfactant coverage.


\section*{Conclusions}
We have shown that appropriately designed long DNA molecules can act as surfactants in an NS droplet system, with the ability to slow coalescence (and thus limit droplet size) and to prevent adhesion to an external solid surface. The surfactants are capable of generating droplets with tunable sizes ranging from $10 \mu$m to $30$ nm. We attribute this effect to the entropic disjoining pressure that the DNA surfactant layer imparts on two droplets that approach each other. We further found that the surfactant prevents the droplets from adhering to hydrophobic and BSA coated surfaces. We suggest that the size control and anti-adhesive properties of the system could give insight into size control mechanisms of biological condensates, or could open new directions in controlling NS synthetic cells. Further, we note that the smallest particles created are similar in size ($\approx$ 10-100 nm) to lipid vesicle therapeutics\cite{hou2021}, such as the RNA-based COVID vaccines, which might indicate the ability to use the methods described here to create nanoscale nucleic-acid drug delivery particles.

It is intriguing that this study found relatively strong effects on droplet size and adhesion using a DNA molecule that is only moderately surface active-- for example, our estimates indicate that, at a 10:1 ratio of NS to surfactant, the DNA surfactant creates a relatively sparse brush layer, and is not strongly depleted from the droplet interior. It is likely related to this that some of the observed effects only occurred at relatively high concentrations of the surfactant. We expect that surface density could be increased by increasing surfactant-NS binding, but this would also increase the presence of long DNA  inside the droplet. There is thus likely an optimum design that maximizes surfactant functionality while minimizing its presence in non-interfacial locations. Future work could attempt to find this optimum by investigating molecular design changes, such as altered sizes, geometries, or binding strengths of the NSs and/or surfactants, and so permit control of droplet behavior with small amounts of added surfactant. 


\section*{Methods}

\subsection*{NS synthesis} To create NSs, 4 single-stranded DNA oligos (See Table S1 for sequences) were mixed, each at a concentration of $50~\mathrm{\mu M}$, in a solution with a total volume of $50~\mathrm{\mu L}$ in $10~\mathrm{mM}$ tris-HCl buffer solution. The mixture was heated to $95^{\circ}$C for $10$ minutes and cooled to $4^{\circ}$C at a constant rate of $0.5^{\circ}$ C/min in a thermocycler. Proper NS assembly was verified through agarose gel electrophoresis. Binding and phase-separation between annealed NSs was activated upon addition of $1$ M NaCl.

\subsection*{DNA surfactant synthesis} DNA surfactants were created using auto-sticky PCR\cite{Gal1999}, using Taq polymerase and lambda phage DNA as the template. Specifically, the forward PCR primer consisted of a template-binding domain that is separated by an abasic site from a 7-base sticky end; the polymerase is arrested by the abasic site, thus the PCR procedure generated double-stranded DNA labeled with the 7-base sticky end. See Table S2 for primer sequences. PCR products were purified using Zymo DNA Clean \& Concentrator-25, and concentrated using Amicon Ultra Centrifugal Filters.

\subsection*{Droplet preparation} The NS and DNA surfactant solutions were mixed together, at the indicated concentrations, in 1 M NaCl. The mixture was heated to 50$^\circ$~C for 30 minutes to melt all sticky ends and homogenize the solution, then the samples were placed on a rotor for 1-2 hours at room temperature to form droplets prior to imaging.

\subsection*{Flow cell preparation} For adhesion tests on hydrophilic glass, cover slips were cleaned by immersion in $2\%$ Hellmanex solution and $10$ minute sonication, rinsed with DI water, then submerged in a $10$ mg/mL BSA protein solution for at least $24$ hours. The cover slips were then rinsed with DI water and dried. Hydrophobic glass was prepared by rinsing cover slips with acetone, isopropanol, and DI water, then subjecting the cover slip to a plasma clean. The cover slips were then coated by placing several drops of Sigmacote (a silanizing reagent) on top, incubating for 30 s, then rinsing with ethanol and air drying. Flow cells were assembled by sandwiching parafilm as a spacer in between the treated cover slip and a top cover slip. 

\subsection*{Fluorescence microscopy and analysis} Confocal images (Fig.~\ref{fig:figure1}) were acquired with a Leica SP8 resonant scanning microscope, using droplets that were fluorescently labeled by tagging 5\% of NSs with Cy5 and 100\% of the DNA surfactant with Cy3.  Epi-fluorescent images (Figs.~\ref{fig:figure2} and~\ref{fig:figure4}) were acquired with a Nikon Ti2-E, using droplets that were fluorescently labeled by tagging 5\% of the NSs with Cy3 and using unlabeled surfactant. Droplet sizes were calculated from images using the algorithm of Crocker and Grier \cite{Crocker1996}. Droplet radii measured to be smaller than one pixel ($0.18~\mathrm{\mu m}$) were discarded from the analysis.

\subsection*{Dynamic Light Scattering} DLS was carried out using a DynaPro NanoStar from Wyatt Technology, with correlation data analyzed using Wyatt DYNAMICS software. Each $40~\mu$L sample, with the indicated concentration of NS and DNA surfactant (both of which were unlabeled), was transferred into an Eppendorf UVette disposable cuvette, and measured twice. The size distribution for each measurement was calculated from averaging correlation functions from a total of $20$ trials with a correlation time of $5$ seconds for each trial.

\subsection*{TEM imaging} The samples with 0.4~$\mu$M unlabeled NS and 0.4~$\mu$M unlabeled surfactant were imaged using a JEM-1011 transmission electron microscope (JEOL) at $80$ kV. The DNA samples were adsorbed ($2~\mu$L, $3$ min) on plasma-exposed (Argon, $60$ s) carbon-coated grids (Formvar, Cu, Ted Pella Inc.) and then negatively stained with $1\%$ uranyl formate (UF). For this, the grid was dabbed with filter paper and then dipped into a $5~\mu$L droplet of UF deposited on Parafilm, followed by dabbing again. The grid was then dipped into a second $5~\mu$L droplet of UF for $8$ s, followed again by dabbing against a filter paper. The grids were left to dry for $10$ min before imaging.

\section*{Author Contributions}
O.S. and S.W. conceived the idea. D.G. primarily carried out the experiments, G.R.A. assisted in sample preparation, A.N. acquired confocal images, and T.L. acquired electron micrographs. S.W. developed the MATLAB code for droplet size measurement. O.S. and D.G. wrote the manuscript, with input from all other authors.

\section*{Conflicts of interest}
O.S., S.W, and D.G. are inventors in a patent application associated with this work.

\section*{Acknowledgements}
We thank Ben Lopez of the NRI-MCDB microscopy facility for assistance in obtaining the confocal PSF. This project was sponsored by the W. M. Keck Foundation under Award No. SB200139. We acknowledge the use of the UCSB NRI-MCDB Microscopy Facility and the Resonant Scanning Confocal supported by NSF MRI grant 1625770; the Biological Nanostructures Laboratory within the California NanoSystems Institute, supported by the University of California, Santa Barbara and the University of California, Office of the President; and the MRL Shared Experimental Facilities, supported by the MRSEC Program of the NSF under Award No. DMR 2308708, a member of the NSF-funded Materials Research Facilities Network. 



\balance


\bibliographystyle{rsc} 

\providecommand*{\mcitethebibliography}{\thebibliography}
\csname @ifundefined\endcsname{endmcitethebibliography}
{\let\endmcitethebibliography\endthebibliography}{}

\end{document}